\documentclass[12pt,a4paper]{article}
\usepackage[top=25truemm,bottom=35truemm,left=25truemm,right=25truemm]{geometry}
\usepackage{amsmath,amssymb}
\usepackage[dvipdfmx]{graphicx}

\begin{document}
\setlength{\baselineskip}{18pt}
\begin{titlepage}
\begin{flushright}
\begin{tabular}{l}
 SU-HET-01-2016\\
\end{tabular} 
\end{flushright}

\vspace*{1.2cm}
\begin{center}
{\Large \bf 
Vacuum stability and naturalness in type-II seesaw
}
\end{center}
\lineskip .75em
\vskip 1.5cm

\begin{center}
{\large Naoyuki Haba}$^1$,
{\large Hiroyuki Ishida}$^1$,
{\large Nobuchika Okada}$^2$,\\ and
{\large Yuya Yamaguchi}$^{1,3}$\\

\vspace{1cm}

$^1${\it Graduate School of Science and Engineering, Shimane University,\\
 Matsue 690-8504, Japan}\\
$^2${\it Department of Physics and Astronomy, University of Alabama,\\
 Tuscaloosa, Alabama 35487, USA}\\
$^3${\it Department of Physics, Faculty of Science, Hokkaido University,\\
 Sapporo 060-0810, Japan}\\

\vspace{10mm}
{\bf Abstract}\\[5mm]
{\parbox{14.5cm}{\hspace{5mm}
%%%%%%%%%%%%%%%%%%%%%%%%%%%%%%%%%%%%%
%             ABSTRACT                      %
%%%%%%%%%%%%%%%%%%%%%%%%%%%%%%%%%%%%%
We study the vacuum stability and perturbativity conditions in the minimal type-II seesaw model.
These conditions give characteristic constraints to the model parameters.
In the model,
 there is a $SU(2)_L$ triplet scalar field,
 which could cause a large Higgs mass correction.
From the naturalness point of view,
 heavy Higgs masses should be lower than $350\,{\rm GeV}$,
 which may be testable by the LHC Run-II results.
Due to the effects of the triplet scalar field,
 the branching ratios of the Higgs decay ($h\to \gamma \gamma, Z\gamma$)
 deviate from the standard model,
 and a large parameter region is excluded
 by the recent ATLAS and CMS combined analysis of $h\to \gamma \gamma$.
Our result of the signal strength for $h\to \gamma \gamma$ is $R_{\gamma \gamma} \lesssim 1.1$,
 but its deviation is too small to observe at the LHC experiment.

}}
\end{center}
\end{titlepage}

%%%%%%%%%%%%%%%%%%%%%%%%%%%%%%%%%%%%%%%%%%%%%
\section{Introduction} \label{intro}
%%%%%%%%%%%%%%%%%%%%%%%%%%%%%%%%%%%%%%%%%%%%%%%%

Current experimental results at the LHC are almost consistent with the predictions in the standard model (SM).
However, 
 the discovery of neutrino oscillations established that active neutrinos are massive,
 and their masses are much smaller than those of the other SM fermions.
Since the SM cannot explain nonzero neutrino masses,
 the existence of nonzero neutrino masses is evidence of physics beyond the SM.
The simplest way to obtain nonzero neutrino masses is breaking the global $B-L$ symmetry,
 which is expressed by an effective dimension-5 operator~\cite{Weinberg:1979sa}.
There are three ways to induce the effective dimension-5 operator at tree level,
 that is, the so-called seesaw mechanism.
There are additional particles to the SM:
 the SM gauge singlet Majorana neutrinos,
 an $SU(2)_L$ triplet scalar field with hypercharge $Y=2$,
 and an $SU(2)_L$ triplet fermion with hypercharge $Y=0$
 in type-I~\cite{seesaw}, II~\cite{Magg:1980ut}-\cite{Mohapatra:1980yp}, and III~\cite{Foot:1988aq} seesaw mechanism, respectively.
Their collider phenomenologies have been studied in Ref.~\cite{Deppisch:2015qwa} for example.

In this paper, we will focus on the minimal type-II seesaw model with a single $SU(2)_L$ triplet scalar field.
The existence of the triplet scalar field can significantly change the electroweak (EW) vacuum structure,
 and the vacuum can become stable up to the Planck scale~\cite{Gogoladze:2008gf}-\cite{Bonilla:2015eha}.
The vacuum stability and perturbativity conditions yield characteristic constraints between model parameters.
%  and we will find that allowed parameter space is much smaller
%  than that obtained by unitarity conditions in stead of the perturbativity condition.
In addition, since the triplet scalar field couples directly to the SM gauge bosons ($W^\pm$, $Z$, $\gamma$),
 its VEV affects $\rho$-parameter at tree level,
 and decay rates of the SM-like Higgs boson ($h\to \gamma \gamma, Z\gamma$)
 are different from the SM case.
Thus, the type-II seesaw model is relatively easy to test at the collider experiments
 compared to type-I and III seesaw models.
We will find that large parameter region can be excluded
 by the recent ATLAS and CMS combined analysis for the signal strength of $h\to \gamma \gamma$.

On the other hand,
 the gauge hierarchy problem generally arises
 when the SM is extended with some heavy particles which couple to the Higgs doublet.
In the SM,
 all operators are renormalizable,
 and there is no quadratic divergence in terms of the dimensional regularization.
Moreover, radiative corrections (or renormalization group evolution) of the Higgs mass term
 does not change its order of magnitude,
 and hence, the SM itself is ${\it natural}$.
However,
 adding a heavy particle into the SM and integrating it out,
 the Higgs mass term receives a contribution of the heavy particle.
It is proportional to $M^2$,
 where $M$ is a heavy particle mass.
When the contribution is much larger than the EW scale,
 there should be a fine-tuning to realize the Higgs mass of $125\,{\rm GeV}$,
 unless the Higgs mass term is protected by a symmetry,
 for example, supersymmetry, shift symmetry, or conformal symmetry (scale invariance).
In this paper, we do not consider such a symmetry,
 but simply impose a naturalness condition
 that contributions of the heavy triplet scalar field should be lower than the measured Higgs mass.
As a result, we will find an upper bound on heavy Higgs masses to be around $350\,{\rm GeV}$.

This paper is organized as follows.
In Sec.~\ref{sec:review},
 we briefly review the type-II seesaw model
 and derive mass eigenstates of the scalar sector.
In Sec.~\ref{sec:stability},
 we summarize the vacuum stability and unitarity conditions,
 and define our naturalness condition.
In Sec.~\ref{sec:result},
 we show the allowed parameter space of scalar quartic couplings
 and branching ratios of $h\to \gamma \gamma, Z\gamma$.
Our conclusions are given in Sec.~\ref{sec:conclusion}.

%%%%%%%%%%%%%%%%%%%%%%%%%%%%%%%%%%%%%%%%%%%%%
\section{Review of the type-II seesaw model} \label{sec:review}
%%%%%%%%%%%%%%%%%%%%%%%%%%%%%%%%%%%%%%%%%%%%%%%%
We consider the minimal type-II seesaw model (for more detailed discussion, see e.g.,~\cite{Accomando:2006ga}),
 where, in addition to the SM fields, a triplet scalar field $\Delta$ is introduced,
 which transforms as $({\bf 1},{\bf 3},2)$ under the $SU(3)_C \times SU(2)_L \times U(1)_Y$ gauge group:
\begin{eqnarray}
	\Delta = \frac{\sigma^i}{\sqrt 2}\Delta_i 
		= \left( \begin{array}{cc}
			\delta^+/\sqrt 2 & \delta^{++}\\
			\delta^0 & -\delta^+/\sqrt 2
		\end{array} \right),
\end{eqnarray}
%$\sigma^i$'s being the usual Pauli matrices, and 
 with $\Delta_1=(\delta^{++}+\delta^0)/\sqrt 2,~\Delta_2=i(\delta^{++}-\delta^0)/\sqrt 2,~\Delta_3=\delta^+$.
The Lagrangian for this model is given by 
\begin{eqnarray}
	{\cal L} = {\cal L}_{\rm kinetic} + {\cal L}_Y - {\cal V}(\Phi,\Delta),
\label{lag}
\end{eqnarray}
 where the relevant kinetic and Yukawa interaction terms are, respectively,
\begin{eqnarray}
	{\cal L}_{\rm kinetic} &=& {\cal L}_{\rm kinetic}^{\rm SM}
		+ {\rm Tr}\left[\left(D_\mu \Delta\right)^\dag \left(D^\mu\Delta\right)\right]\, ,\\
	{\cal L}_Y &=& {\cal L}_Y^{\rm SM}
		- \frac{1}{\sqrt 2}\left(Y_\Delta\right)_{ij} L_i^{\sf T}Ci\sigma_2\Delta L_j+{\rm H.c.}\, .
\end{eqnarray}
Here $C$ is the Dirac charge conjugation matrix with respect to the Lorentz group,
 and 
% $L_i=(\nu_i,\ell_i)_L^{\sf T}$ (with $i=e,\mu,\tau$) is the $SU(2)_L$ lepton doublet, $\Phi=(\phi^+,\phi^0)^{\sf T}$ is the SM Higgs doublet,
\begin{eqnarray} 
	D_\mu \Delta = \partial_\mu \Delta + i\frac{g}{2}[\sigma^a W_\mu^a,\Delta]
		+ i\frac{g'}{2}B_\mu \Delta \qquad (a=1,2,3)
\end{eqnarray}
 is the covariant derivative of the scalar triplet field,
 with the GUT-normalization for the electroweak couplings $g=g_2$ and $g'=\sqrt{3/5}g_1$.

Following the notation of~\cite{Schmidt:2007nq},
 we write the scalar potential in Eq.~(\ref{lag}) as\footnote{
The general form of the potential given in~\cite{Arhrib:2011uy} can be recovered
 with a simple redefinition of the couplings:
 $\lambda\to \lambda/2, ~(\lambda_1+\lambda_2)\to 2\lambda_2, ~\lambda_2\to -2\lambda_3, ~(\lambda_4+\lambda_5)\to \lambda_1, ~\lambda_5\to -\lambda_4/2$,
 and using the identity 
$(\Phi^\dag \Phi){\rm Tr}(\Delta^\dag \Delta)=\Phi^\dag\{\Delta^\dag,\Delta\}\Phi$,
 which is valid for any traceless $2\times 2$ matrix $\Delta$.}  
\begin{eqnarray}
	{\cal V}(\Phi, \Delta) &=& - m_\Phi^2 \Phi^\dagger \Phi + \frac{\lambda}{2} (\Phi^\dagger \Phi)^2
		+ M_\Delta^2 {\rm Tr}(\Delta^\dagger \Delta)
		+ \frac{\lambda_1}{2} \left[ {\rm Tr}(\Delta^\dagger \Delta) \right]^2 \nonumber \\
	&& + \frac{\lambda_2}{2} \left( \left[ {\rm Tr}(\Delta^\dagger \Delta) \right]^2
			- {\rm Tr} \left[ (\Delta^\dagger \Delta)^2 \right] \right)
		+ \lambda_4 (\Phi^\dagger \Phi) {\rm Tr}(\Delta^\dagger \Delta)
		+ \lambda_5 \Phi^\dagger [\Delta^\dagger, \Delta] \Phi \nonumber \\
	&& + \left( \frac{\Lambda_6}{\sqrt{2}} \Phi^{\rm T} i \sigma_2 \Delta^\dagger \Phi + {\rm H.c}. \right).
\label{potential}
\end{eqnarray}
We have chosen $m_\Phi^2>0$ in order to ensure the spontaneous EW symmetry breaking.
Stationary conditions of the scalar potential lead to
\begin{eqnarray}
	m_\Phi^2 &=& \frac{1}{2}\lambda v^2 - \Lambda_6 v_\Delta
		+ \frac{1}{2} (\lambda_4-\lambda_5) v_\Delta^2, \\
	M_\Delta^2 &=& \frac{1}{2} \frac{\Lambda_6 v^2}{v_\Delta}
		- \frac{1}{2} (\lambda_4-\lambda_5) v^2 - \frac{1}{2} \lambda_1 v_\Delta^2,
\label{M_Delta}
\end{eqnarray}
 where $v$ and $v_\Delta$ are VEVs of neutral components of $\Phi$ and $\Delta$, respectively.
The nonzero $v_\Delta$ makes the $\rho$-parameter deviate from unity at the tree level as
\begin{eqnarray}
	\rho \equiv \frac{M_W^2}{M_Z^2 \cos^2 \theta_W}
		= \frac{1+\frac{2v_\Delta^2}{v^2}}{1+\frac{4v_\Delta^2}{v^2}}, 
\end{eqnarray}
 with the gauge boson masses given by
\begin{eqnarray}
	M_W^2 = \frac{g_2^2}{2} \left( v^2 + 2 v_\Delta^2 \right),\qquad
	M_Z^2 = \frac{g_2^2}{2 \cos^2 \theta_W} \left( v^2 + 4 v_\Delta^2 \right),
\end{eqnarray}
 where $g_2$ is the $SU(2)_L$ gauge coupling constant
 and $\theta_W$ is the Weinberg angle.
From the experimental bound $\rho=1.0004^{+0.0003}_{-0.0004}$~\cite{Beringer:1900zz},
 $v_\Delta$ is strongly restricted by
\begin{eqnarray}
	\frac{v_\Delta}{v} \lesssim 0.02,\quad {\rm or\ equivalently}\quad v_\Delta \lesssim 5\,{\rm GeV}.
\end{eqnarray}

In the limit $v_\Delta \ll v$, we obtain from Eq.\,(\ref{M_Delta})
\begin{eqnarray}
	v_\Delta \approx \frac{\lambda_6 M_\Delta v^2}{2 M_\Delta^2 + v^2(\lambda_4- \lambda_5)},
\end{eqnarray}
 where we have defined $\lambda_6 \equiv \Lambda_6 /M_\Delta$.
Then the neutrino mass matrix is given by
\begin{eqnarray}
	(M_\nu)_{ij} = v_\Delta (Y_\Delta)_{ij}
		\approx \frac{\lambda_6 M_\Delta v^2}{2 M_\Delta^2 + v^2(\lambda_4- \lambda_5)} (Y_\Delta)_{ij},
\end{eqnarray}
 where $i,j = 1, 2, 3$ are flavor indices.
On the other hand, $M_\nu$ is written using Pontecorvo-Maki-Nakagawa-Sakata (PMNS) matrix
 $U$~\cite{Maki:1962mu,Pontecorvo:1967fh}
 as $M_\nu = U^{\rm T} M_\nu^{\rm diag} U$,
 where $M_\nu^{\rm diag} = {\rm diag}(m_1, m_2, m_3)$ is the diagonal neutrino mass eigenvalue matrix.
Using the central values of a recent neutrino oscillation data~\cite{Agashe:2014kda},
 the order of magnitude of neutrino Yukawa coupling matrix is estimated as
\begin{eqnarray}
	Y_\Delta = \frac{10^{-2}\,{\rm eV}}{v_\Delta} \times \mathcal{O}(1)_{3\times 3},
\end{eqnarray}
 where the last $3 \times 3$ matrix can be calculated by mass eigenvalues, mixing angles, Dirac CP phase,
 and two Majorana phases.
Since the Yukawa coupling should be less than unity for perturbation theory,
 $v_\Delta$ is bounded from below as $v_\Delta \gtrsim \mathcal{O}(10^{-2}\,{\rm eV})$.

%%%%%%%%%%%%%%%%%%%%%%%%%%%%%%%%%%%%
% \section{Scalar masses and mixings}
%%%%%%%%%%%%%%%%%%%%%%%%%%%%%%%%%%%%

In the rest of this section,
 we explain masses and mixings of the scalar fields.
Expanding the scalar fields $\phi^0$ and $\delta^0$ around their VEVs
 ($\phi^0=(v+\phi+i\chi)/\sqrt{2}$ and $\delta^0=(v_\Delta+\delta+i\eta)/\sqrt{2}$),
 we obtain 10 real-valued field components,
% \begin{eqnarray}
% 	\Phi = \left( \begin{array}{cc}
% 		\phi^+ \\ (v+\phi+i\chi)/\sqrt{2}
% 		\end{array} \right), \qquad
% 	\Delta = \left( \begin{array}{cc}
% 		\delta^+ /\sqrt{2} & \delta^{++}\\
% 		(v_\Delta+\delta+i\eta)/\sqrt{2} & -\delta^+ /\sqrt{2}
% 		\end{array} \right),
% \end{eqnarray}
 which yields a $10\times 10$ squared mass matrix for the scalars.
%  upon minimization of the scalar potential ${\cal V}(\Phi,\Delta)$ in Eq.~(\ref{potential}) with respect to the VEVs.
There are seven physical massive eigenstates $H^{\pm\pm},H^\pm,h,H^0,A^0$
 and three massless Goldstone bosons $G^\pm,G^0$,
 which are eaten by the SM gauge bosons $W^\pm,Z$.
The physical mass eigenvalues for the scalar sector are given as follows:
\begin{eqnarray}
	M^2_{H^{\pm\pm}} &=& M_\Delta^2+\frac{1}{2}(\lambda_4+\lambda_5)v^2
		+\frac{1}{2}(\lambda_1+\lambda_2)v_\Delta^2, \label{doub-mass}\\
	M^2_{H^\pm} &=& \left(M_\Delta^2+\frac{1}{2}\lambda_4v^2
		+\frac{1}{2}\lambda_1v_\Delta^2\right)\left(1+\frac{2v^2_\Delta}{v^2}\right),\label{sing-mass}\\
	M^2_{A^0} &=& \left(M_\Delta^2+\frac{1}{2}(\lambda_4-\lambda_5)v^2
		+\frac{1}{2}\lambda_1v_\Delta^2\right)\left(1+\frac{4v^2_\Delta}{v^2}\right),\\
	M^2_h &=& \frac{1}{2}\left(A+C-\sqrt{(A-C)^2+4B^2}\right),\\
	M^2_{H^0} &=& \frac{1}{2}\left(A+C+\sqrt{(A-C)^2+4B^2}\right),\\
{\rm with}~~
A = \lambda v^2,&&
B = -\frac{2v_\Delta}{v}\left(M^2_\Delta+\frac{1}{2}\lambda_1 v_\Delta^2\right),~~
C = M^2_\Delta+\frac{1}{2}(\lambda_4-\lambda_5)v^2+\frac{3}{2}\lambda_1 v_\Delta^2.\nonumber
\end{eqnarray}
Note that among the two $CP$-even neutral Higgs bosons,
 $M_{H^0}>M_{h}$ is satisfied for $A<C$,
 which we will impose in our analysis.

The mixing between the doublet and triplet scalar fields
 in the charged, $CP$-even and $CP$-odd scalar sectors are, respectively, given by 
\begin{eqnarray}
	\left( \begin{array}{c} G^\pm \\ H^\pm \end{array} \right)
		&=& \left( \begin{array}{cc}
			\cos\beta' & \sin\beta'\\
			-\sin\beta' & \cos\beta'
		\end{array} \right)
		\left( \begin{array}{c} \phi^\pm \\ \delta^\pm \end{array} \right),\\
	\left( \begin{array}{c} h \\ H^0 \end{array}\right)
		&=& \left( \begin{array}{cc}
			\cos\alpha & \sin\alpha\\
			-\sin\alpha & \cos\alpha
		\end{array} \right)
		\left( \begin{array}{c} \phi \\ \delta \end{array} \right),\\
	\left(\begin{array}{c} G^0 \\ A^0 \end{array}\right)
		&=& \left( \begin{array}{cc}
			\cos\beta & \sin\beta\\
			-\sin\beta & \cos\beta
		\end{array} \right)
		\left(\begin{array}{c} \chi \\ \eta \end{array} \right),
\end{eqnarray} 
 where the mixing angles are given by%~\cite{arhrib1}
\begin{eqnarray}
	\tan\beta' &=& \frac{\sqrt 2 v_\Delta}{v},\label{mix1}\\
	\tan\beta &=& \frac{2v_\Delta}{v} = \sqrt 2 \tan\beta',\label{mix2}\\
	\tan{2\alpha} &=& \frac{2B}{A-C}
		= \frac{4v_\Delta}{v}\frac{M_\Delta^2+\frac{1}{2}\lambda_1 v_\Delta^2}{M_\Delta^2+\frac{1}{2}(\lambda_4-\lambda_5-2\lambda)v^2+\frac{3}{2}\lambda_1 v_\Delta^2}
\label{mix3}.
\end{eqnarray} 
Thus, in the limit $v_\Delta\ll v$, the mixing between the doublet and triplet scalars is small,
 unless the $CP$-even scalars $h$ and $H^0$ are close to being mass-degenerate.
In this limit, the mass of the (dominantly doublet) lightest $CP$-even scalar
 is simply given by $M_h^2=\lambda v^2$ (as in the SM) independent of the mass scale $M_\Delta$,
 whereas the other (dominantly triplet) scalars have $M_\Delta$-dependent mass.
% The mass scale $M_\Delta$ will be simply referred to as the ``seesaw scale" for the rest of our paper.  

After integrating out the heavy Higgs triplets,
 the effective scalar potential is given by
\begin{eqnarray}
	V_{\rm eff} = - m_\Phi^2 \Phi^\dagger \Phi + \frac{1}{2} (\lambda-\lambda_6^2) (\Phi^\dagger \Phi)^2.
\end{eqnarray}
At $\mu = M_\Delta$, the following matching condition is satisfied:
\begin{eqnarray}
	\lambda_{\rm SM} = \lambda-\lambda_6^2,
\label{shift}
\end{eqnarray}
 where $\lambda_{\rm SM}$ is the SM Higgs quartic coupling.
Note that the EW vacuum can be stable by a sufficiently large $\lambda_6$ as shown later.

%%%%%%%%%%%%%%%%%%%%%%%%%%%%%%%%%%%%
\section{Vacuum stability and naturalness} \label{sec:stability}
%%%%%%%%%%%%%%%%%%%%%%%%%%%%%%%%%%%%

Since $\lambda_{\rm SM}$ becomes negative at around $\mu=10^{8-10}\,{\rm GeV}$ in the SM
 [see~\cite{Buttazzo:2013uya} for example],
 new physics scale, which corresponds to $M_\Delta$ in our case,
 have to appear before $\lambda_{\rm SM}$ becomes negative.
To ensure that the scalar potential (\ref{potential}) is bounded from below,
 the necessary and sufficient conditions are given by~\cite{Arhrib:2011uy}
\begin{eqnarray}
	&& \lambda \geq 0,\quad \lambda_1 \geq 0,\quad 2 \lambda_1 + \lambda_2 \geq 0, \nonumber \\
	&& \lambda_4 + \lambda_5 + \sqrt{\lambda \lambda_1} \geq 0,\quad
	\lambda_4 + \lambda_5 + \sqrt{\lambda \left( \lambda_1 + \frac{\lambda_2}{2} \right)} \geq 0, \nonumber \\
	&& \lambda_4 - \lambda_5 + \sqrt{\lambda \lambda_1} \geq 0,\quad
	\lambda_4 - \lambda_5 + \sqrt{\lambda \left( \lambda_1 + \frac{\lambda_2}{2} \right)} \geq 0. 
\label{condition}
\end{eqnarray}
In fact, corrections of necessary and sufficient conditions have been recently pointed out
 by Eq.~(19) in Ref.~\cite{Bonilla:2015eha}.
The major difference would appear in the ($\lambda_4$, $\lambda_5$) plane,
 that is, the correct conditions can make the allowed parameter region larger than that by Eq.~(\ref{condition}).
Since the difference is not so large and does not change our main result significantly,
 we will consider Eq.~(\ref{condition}) for a relevant vacuum stability condition.

In addition, the tree-level unitarity of the $S$-matrix for elastic scattering imposes the following constraints~\cite{Arhrib:2011uy}:
\begin{eqnarray}
	&&\lambda \leq \frac{8}{3}\pi,\quad
		\lambda_1-\lambda_2\leq 8\pi,\quad
		4\lambda_1+\lambda_2\leq 8\pi,\quad
		2\lambda_1+3\lambda_2\leq 16\pi,\nonumber \\
	&& |\lambda_5|\leq \frac{1}{2}
		{\rm min}\left[\sqrt{(\lambda\pm 8\pi)(\lambda_1-\lambda_2\pm 8\pi)}\right],\nonumber\\
	&&|\lambda_4| \leq \frac{1}{\sqrt 2}
		\sqrt{\left(\lambda-\frac{8}{3}\pi\right)\left(4\lambda_1+\lambda_2-8\pi\right)}.
\end{eqnarray} 
We also impose the perturbativity condition,
 that is, all quartic couplings are less than $4\pi$ up to the Planck scale.
It turns out that the perturbativity condition more strongly constrains the parameter space than the unitarity condition.
The one-loop beta functions of coupling constants are given in the appendix.

In the ordinary type-II seesaw model,
 the Higgs mass correction is given by \cite{Abada:2007ux}
\begin{eqnarray}
	\delta m_h^2 = \frac{1}{16\pi^2} \left[
		3 \lambda_4 \left( \Lambda^2 - M_\Delta^2 \log \frac{\Lambda^2}{M_\Delta^2}\right)
		- 3 \lambda_6^2 M_\Delta^2 \log \frac{\Lambda^2}{M_\Delta^2} \right],
\label{cor_cutoff}
\end{eqnarray}
 where $\Lambda$ is a UV cutoff.
In our analysis, we neglect the quadratic divergent term,
 because it does not appear in the dimensional regularization.
On the other hand,
 the logarithmic correction appears in the scheme-independent form,
 i.e., the coefficient of logarithmic term is the same for any regularization scheme. 
Thus, we consider only the logarithmic terms in Eq.\,(\ref{cor_cutoff}) for a physical correction,
\begin{eqnarray}
	\delta m_h^2 = - \frac{3}{16\pi^2} \left( \lambda_4 + \lambda_6^2 \right)
		M_\Delta^2 \log \frac{M_{\rm Pl}^2}{M_\Delta^2},
\label{correction}
\end{eqnarray}
 where we have set $\Lambda=M_{\rm Pl}=2.4\times10^{18}\,{\rm GeV}$.
We evaluate a fine-tuning level as $|\delta m_h^2|/M_h^2$,
 where $M_h=125\,{\rm GeV}$ is the experimentally observed Higgs boson mass.
We require the fine-tuning level to be less than unity for the naturalness in our analysis.

Here, we mention results from the different naturalness conditions in the literature.
In Ref.~\cite{Farina:2013mla},
 the authors evaluated the Higgs mass correction at two-loop induced by electroweak interactions,
 and they obtained the upper bound of triplet scalar around 200\,GeV in type-II seesaw,
 while they have not considered the corrections due to couplings not related to active neutrino masses
 except for $\lambda_4$ in our notation.
However, we will find $\lambda_6$ is also important for the naturalness condition (\ref{correction})
 to realize the vacuum stability.
On the other hand,
 in Ref.~\cite{Chabab:2015nel},
 the authors obtained $M_{H^\pm}<288\,{\rm GeV}$ and $M_{H^{\pm\pm}}<351\,{\rm GeV}$
 by considering the naturalness condition in terms of the Veltman condition~\cite{Veltman:1980mj},
 which requires a cancellation of quadratic divergences.
Since we neglect the quadratic divergences for an unphysical quantity,
 our method is completely different from that in Ref.~\cite{Chabab:2015nel}.
However, we will find that our result is accidentally almost the same as their results.

%%%%%%%%%%%%%%%%%%%%%%%%%%%%%%%%%%%%
\section{Numerical analysis} \label{sec:result}
%%%%%%%%%%%%%%%%%%%%%%%%%%%%%%%%%%%%

In this section, we show some numerical results with scatter plots,
 which satisfy the vacuum stability condition~(\ref{condition}) and the perturbativity condition.
In our analysis, we solve the renormalization group equations at two-loop level
 with a one-loop threshold correction for $\lambda$,
 and we restrict the regions of some parameters as follows:
\begin{eqnarray}
	&& M_h = 125.09 \pm 0.21 \ ({\rm stat.}) \pm 0.11 \ ({\rm syst.})\ {\rm GeV}~\cite{Aad:2015zhl},\nonumber \\
	&& M_t = 173.34 \pm 0.27 \ ({\rm stat.}) \pm 0.71 \ ({\rm syst.})\ {\rm GeV}~\cite{ATLAS:2014wva}, \nonumber \\
	&& \alpha_s = 0.1184 \pm 0.0007~\cite{Bethke:2012jm}, \nonumber \\
	&& 200\,{\rm GeV} \leq M_\Delta \leq 10^{12}\,{\rm GeV}, \quad
	0.01\,{\rm eV} \leq v_\Delta \leq 5\,{\rm GeV}. %, \nonumber \\
% 	&& \quad 0 \leq \lambda_1 (M_\Delta) \leq 1, \qquad \qquad \quad
% 	-1.5 \leq \lambda_2 (M_\Delta) \leq 1.5, \nonumber \\
% 	&& -0.5 \leq \lambda_4 (M_\Delta) \leq 1, \qquad \qquad \quad
% 	-1 \leq \lambda_5 (M_\Delta) \leq 1.
\end{eqnarray}
Although we take $M_\Delta$ up to $10^{12}\,{\rm GeV}$,
 the following numerical results always satisfy the requirement that
 $M_\Delta$ is lower than the energy scale, at which $\lambda_{\rm SM}$ becomes negative.
The upper and lower bounds of $v_\Delta$ are given by
 $\rho$-parameter bound and naturalness of neutrino Yukawa coupling, respectively,
 as mentioned in Sec.~\ref{sec:review}.
We also adopt the following constraints for the charged Higgs boson masses:
\begin{eqnarray}
	&& M_{H^{\pm\pm}}>550\,{\rm GeV\ for}\ v_\Delta < 10^{-4}\,{\rm GeV},\\
	&& v_\Delta M_{H^{\pm\pm}}>150\,{\rm eV\, GeV},\\
	&& |M_{H^{\pm\pm}}-M_{H^{\pm}}| < 40\,{\rm GeV},
\end{eqnarray}
 which correspond to the experimental bounds on
 decay mode of $H^{\pm\pm}\to \ell^{\pm}\ell^{\pm}$~\cite{ATLAS:2014kca},
 lepton flavor violating decays~\cite{Akeroyd:2009nu,Fukuyama:2009xk},
 and the electroweak precision data~\cite{Chun:2012jw}, respectively.

%%%%%%%%%%%%%%%%%%%%%%%%%
\subsection{Allowed parameter space}
%%%%%%%%%%%%%%%%%%%%%%%%%

\begin{figure}
\begin{center}
	\includegraphics[height=7cm]{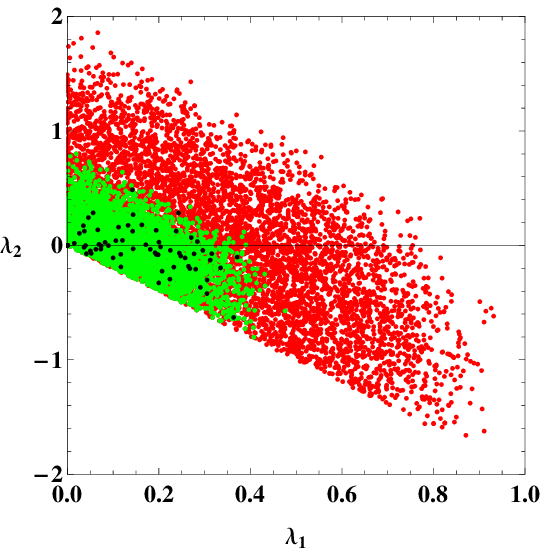} \hspace{5mm}
	\includegraphics[height=7cm]{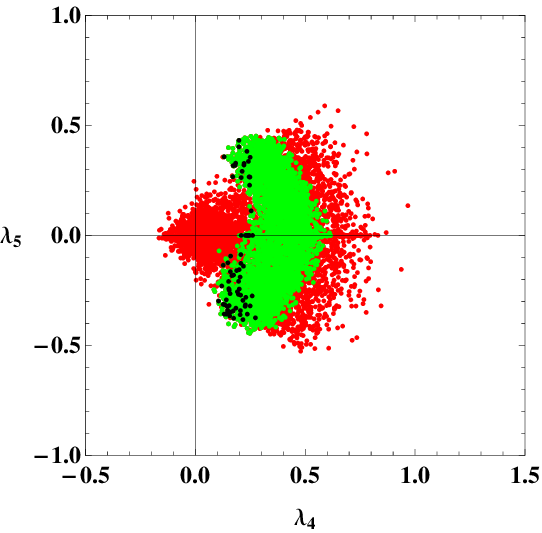}
\end{center}
\caption{Scatter plots in the ($\lambda_1$, $\lambda_2$) plane (left) and the ($\lambda_4$, $\lambda_5$) plane (right),
 which satisfy the vacuum stability and perturbativity conditions.
For $200\,{\rm GeV} \leq M_\Delta \leq 1\,{\rm TeV}$ the allowed parameter space is limited in the green region.
The black dots satisfy the naturalness condition of $|\delta m_h^2|<M_h^2$.}
\label{scatter}
\end{figure}

Figure\,\ref{scatter} shows scatter plots
 in the ($\lambda_1$, $\lambda_2$) plane (left) and the ($\lambda_4$, $\lambda_5$) plane (right),
 which satisfy the vacuum stability and perturbativity conditions.
The green dots in Fig.~\ref{scatter} correspond to the allowed parameter space
 for $200\,{\rm GeV} \leq M_\Delta \leq 1\,{\rm TeV}$,
 in which the new quartic couplings ($\lambda_i$, $i=1,2,4,5$) should be sufficiently small
 to keep the perturbativity up to the Planck scale.
The black dots satisfy $|\delta m_h^2|<M_h^2$.
The lower bound of $\lambda_1$ is $-\lambda_2/2$ because of $2\lambda_1+\lambda_2\geq 0$,
 and the upper bounds of $\lambda_1$ and $\lambda_2$ come from the perturbativity condition.
Since the third terms of the last four inequalities in Eq.~(\ref{condition}) 
 are negligible for a sufficiently large $\lambda_4\pm\lambda_5$,
 the vacuum stability requires $-\lambda_4 \lesssim \lambda_5 \lesssim \lambda_4$.
Note that, when both $|\lambda_4|$ and $|\lambda_5|$ are small,
 the vacuum can become stable only by a sufficiently large $\lambda_6$,
 which will be explained in detail below.
The allowed parameter space shown in Fig.\,\ref{scatter} is much smaller than that in Ref.~\cite{Dev:2013ff},
 in which only the unitarity condition has been considered.

\begin{figure}
\begin{center}
	\includegraphics[height=7cm]{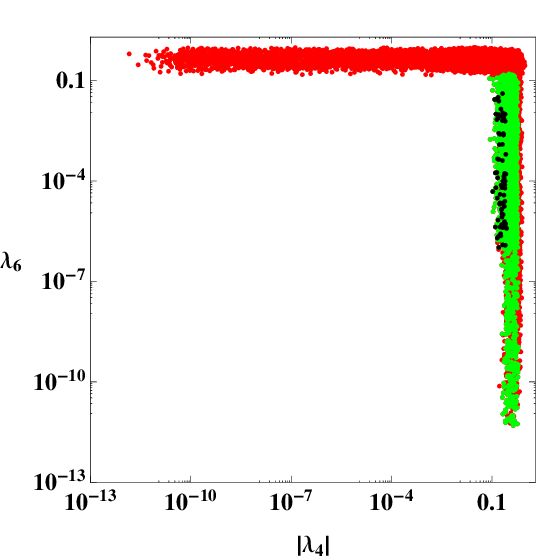} \hspace{5mm}
	\includegraphics[height=7cm]{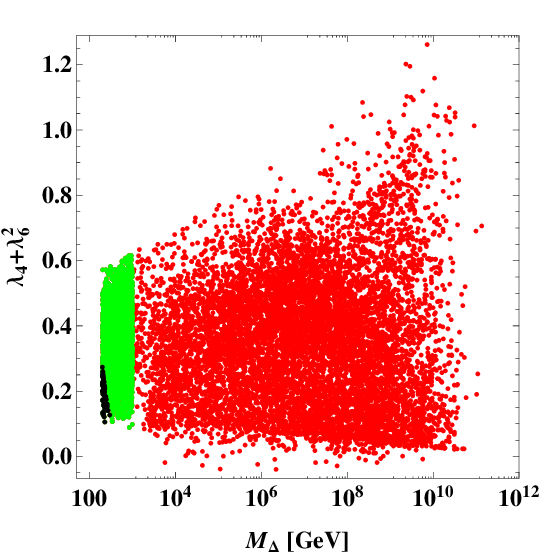}
\end{center}
\caption{Scatter plots in the ($|\lambda_4|$, $\lambda_6$) plane (left)
 and the ($M_\Delta$ [GeV], $\lambda_4 + \lambda_6^2$) plane (right),
 which satisfy the vacuum stability and perturbativity conditions.
The setup is the same as in Fig.~\ref{scatter}.}
\label{lambda4-6}
\end{figure}

The left panel of Fig.~\ref{lambda4-6} shows scatter plots in the ($|\lambda_4|$, $\lambda_6$) plane.
There is no allowed parameter space for $|\lambda_4| < 0.01$ and $\lambda_6<0.01$.
Then we find $\lambda_4+\lambda_6^2 = {\cal O}(1)$ for almost all values of $M_\Delta$,
 which is shown in the right panel of Fig.~\ref{lambda4-6}.
If both $|\lambda_4|$ and $\lambda_6$ are sufficiently small,
 the Higgs mass correction would be smaller than the Higgs mass.
However, the vacuum stability cannot be realized by such small parameters.

\begin{figure}
\begin{center}
	\includegraphics[height=6cm]{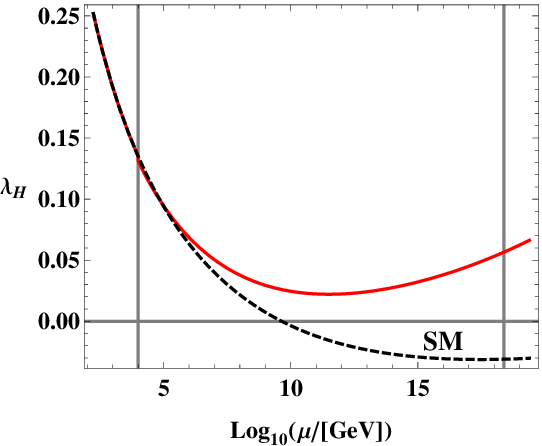} \hspace{5mm}
	\includegraphics[height=6cm]{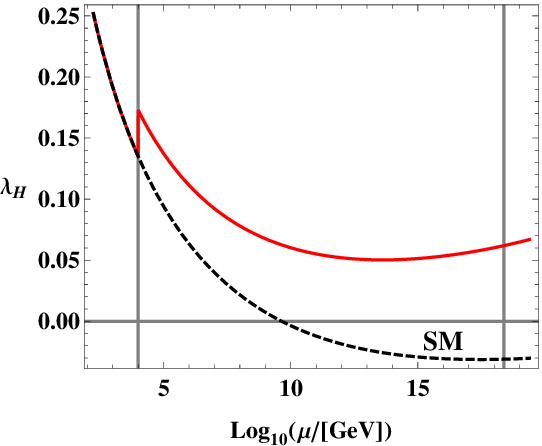}
\end{center}
\caption{Running of $\lambda_H
 (=\lambda_{\rm SM}\ {\rm for\ \mu < M_\Delta}, \lambda\ {\rm for\ \mu \geq M_\Delta})$.
We have taken $\lambda_4=0.2$ and $\lambda_6=3\times 10^{-4}$ in the left panel,
 while $\lambda_4=0.1$ and $\lambda_6=0.2$ in the right panel.
Other quartic couplings are commonly taken as $\lambda_1=\lambda_2=\lambda_5=0.1$.
The black dashed line corresponds to the SM case.
The vertical lines show $M_\Delta=10\,{\rm TeV}$ and $M_{\rm Pl}$, respectively.
}
\label{running}
\end{figure}

To stabilize the EW vacuum, there are two types of solutions.
We show the running of $\lambda_H
 (=\lambda_{\rm SM}\ {\rm for\ \mu < M_\Delta}, \lambda\ {\rm for\ \mu \geq M_\Delta})$
 for typical input parameters in Fig.~\ref{running}.
The left panel of Fig.~\ref{running} corresponds to the large positive contribution to a beta function of $\lambda$ ($\beta_\lambda$),
 which is realized by a large $|\lambda_4|$ and/or $|\lambda_5|$ [see Eq.\,(\ref{beta})].
The right panel of Fig.~\ref{running} corresponds to a discontinuous shift between $\lambda_{\rm SM}$ and $\lambda$,
 which is realized by a large $\lambda_6$ [see Eq.\,(\ref{shift})].
When $|\lambda_4|$ and $\lambda_6$ are sufficiently small but $|\lambda_5|$ is sufficiently large,
 the EW vacuum likely become stable,
 because $\lambda_5$ also positively contributes to $\beta_\lambda$.
In that case, however, the last two conditions in Eq.\,(\ref{condition}) cannot be satisfied.
Thus, when the EW vacuum becomes stable,
 the Higgs mass correction usually becomes larger than the Higgs mass.

\begin{figure}
\begin{center}
	\includegraphics[height=6.2cm]{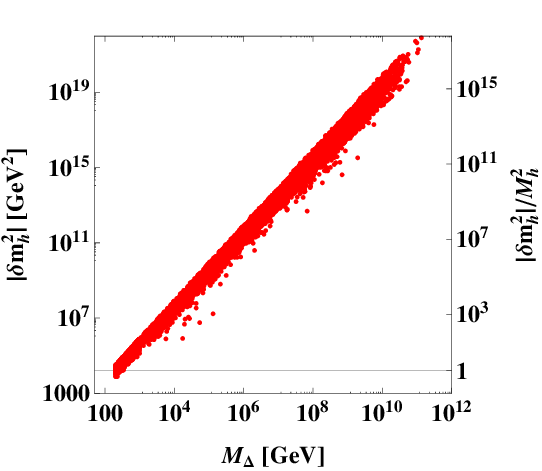} \hspace{2mm}
	\includegraphics[height=6.2cm]{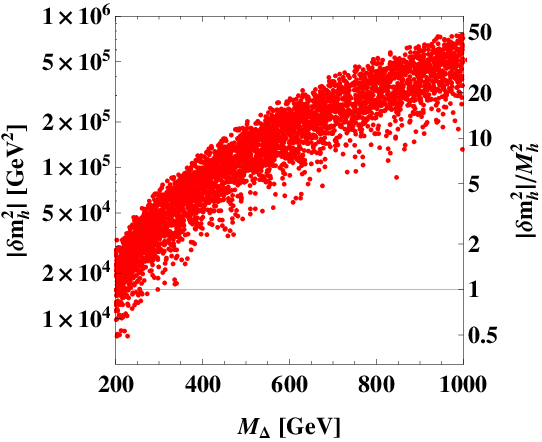}
\end{center}
\caption{$M_\Delta$ dependence of the Higgs mass correction for all allowed region of $M_\Delta$ (left)
 and $200\,{\rm GeV} \leq M_\Delta \leq 1\,{\rm TeV}$ (right).
The horizontal line shows the Higgs mass squared, i.e., $M_h^2 = (125\,{\rm GeV})^2$.}
\label{fig:mass}
\end{figure}

Figure~\ref{fig:mass} shows $M_\Delta$ dependence of the Higgs mass correction,
 which is calculated by Eq.\,(\ref{correction}).
From the right panel of Fig.~\ref{fig:mass},
 we find that the naturalness condition requires $M_\Delta \lesssim 350\,{\rm GeV}$.
Below this bound, the minimal type-II seesaw model can be testable by the LHC Run-II results~\cite{Han:2015hba}
 [also see Refs.~\cite{Akeroyd:2005gt}-\cite{Chun:2013vma} for the LHC phenomenology]. 
In the case of $M_{H^{\pm \pm}}<M_{H^\pm}<M_{H^0/A^0}$,
 from four-lepton signal at the 14\,TeV LHC experiment with $300\,{\rm fb}^{-1}$,
 we can potentially probe up to a mass $M_{H^{\pm \pm}}\sim 600\ (700)\,{\rm GeV}$
 for the normal (inverted) hierarchy of active neutrino masses.
In the case of $M_{H^{\pm \pm}}>M_{H^\pm}>M_{H^0/A^0}$,
 with an integrated luminosity $\sim 500\,{\rm fb}^{-1}$, 
 the triplet scalars can be fully reconstructed at the 14\,TeV LHC.

%%%%%%%%%%%%%%%%%%%%%%%%%
\subsection{Predictions for the decay rate of $h \rightarrow \gamma \gamma$, $Z\gamma$}
%%%%%%%%%%%%%%%%%%%%%%%%%

In the SM, the decay $h\to \gamma \gamma (Z\gamma)$ at the one-loop level is mediated by
 the virtual exchange of SM fermions (dominantly the top-quark) and the $W$-boson.
In the type-II seesaw model,
 there are additional contributions from the new charged Higgs bosons~\cite{Arhrib:2011uy}.\footnote{
There are also other contributions in extended type-II seesaw models,
 for example, in Refs.~\cite{Wang:2012ts,Chun:2013ft}.
} 
The decay rates of $h\rightarrow \gamma\gamma$ is given by~\cite{Chen:2013vi,Chen:2013dh}
\begin{eqnarray}
	\Gamma(h\rightarrow \gamma\gamma) &=& \frac{\alpha^2 G_F M_h^3}{128\sqrt{2}\pi^3} \bigg|
		\sum_f N_c Q_f^2 g_{hf\bar{f}}A^{\gamma\gamma}_{1/2}(\tau_h^f)
		+ g_{hW^+W^-} A^{\gamma\gamma}_1(\tau_h^W) \nonumber\\
		&& + \tilde{g}_{hH^\pm H^\mp} A^{\gamma\gamma}_0(\tau_h^{H^\pm})
		+ 4 \tilde{g}_{hH^{\pm\pm} H^{\mp\mp}} A^{\gamma\gamma}_0(\tau_h^{H^{\pm\pm}}) \bigg|^2\,,
\label{hgg}
\end{eqnarray}
 where  $\alpha$ is the fine-structure constant,
 $G_F$ is the Fermi coupling constant,
 $N_c=3 (1)$ for quarks (leptons),
 and $Q_f$ is the electric charge of the fermion in the loop.
In the same way, the decay rate of $h\rightarrow Z\gamma$ is given by
\begin{eqnarray}
	\Gamma(h\to Z\gamma) &=& \frac{\alpha^2 M_h^3}{128\pi^3 v^2}
		\left(1-\frac{M_Z^2}{M_h^2}\right)^3 \bigg|
		\frac{1}{s_W c_W} \sum_f N_c Q_f (2I_3^f-4Q_fs_W^2) A_{1/2}^{Z\gamma}(\tau_h^f,\tau_Z^f)
		\nonumber\\
		&& + \cot\theta_W g_{hW^+W^-} A_1^{Z\gamma}(\tau_h^W,\tau_Z^W)
		- 2 g_{ZH^\pm H^\mp} \tilde{g}_{hH^\pm H^\mp}
		A_0^{Z\gamma}(\tau_h^{H^\pm},\tau_Z^{H^\pm}) \nonumber\\
		&& - 4 g_{ZH^{\pm\pm}H^{\mp\mp}} \tilde{g}_{hH^{\pm\pm}H^{\mp\mp}}
		A_0^{Z\gamma}(\tau_h^{H^{\pm\pm}},\tau_Z^{H^{\pm\pm}})\bigg|^2\,,
\label{hZg}
\end{eqnarray}
 where $s_W\equiv \sin\theta_W$, $c_W\equiv \cos\theta_W$,
 $\tau_h^i=4M_i^2/M_h^2$, $\tau_Z^i=4M_i^2/M_Z^2$ ($i=f,W,H^\pm,H^{\pm\pm}$),
 and $I_3^{t,s}$ are the third isospin components of the fermion.
In these equations,
 the first two terms in the squared amplitude are the SM fermion and $W$-boson contributions, respectively,
 whereas the last two terms correspond to the $H^\pm$ and $H^{\pm\pm}$ contributions.
We consider only the top quark contribution for the SM fermion,
 because the other fermion contributions are negligible.
The relevant loop functions are defined as
\begin{eqnarray}
	A_0^{\gamma\gamma} (x) &=& -x^2[x^{-1}-f(x^{-1})]\,, \nonumber\\
	A_{1/2}^{\gamma\gamma} (x) &=& 2x^2[x^{-1}+(x^{-1}-1)f(x^{-1})]\,, \nonumber\\
	A_1^{\gamma\gamma}(x) &=& -x^2[2x^{-2}+3x^{-1}+3(2x^{-1}-1)f(x^{-1})]\,, \nonumber\\
	A_0^{Z\gamma}(x,y) &=& I_1(x,y)\,, \\
	A_{1/2}^{Z\gamma} (x,y) &=& I_1(x,y)-I_2(x,y)\,, \nonumber\\
	A_1^{Z\gamma}(x,y) &=& 4(3-\tan^2\theta_W)I_2(x,y)
		+ [(1+2x^{-1})\tan^2 \theta_W-(5+2x^{-1})]I_1(x,y)\,, \nonumber
\end{eqnarray}
 where
\begin{eqnarray}
	I_1(x,y) &=& \frac{x y}{2(x-y)} + \frac{x^2 y^2}{2(x-y)^2}[f(x^{-1})-f(y^{-1})]
		+ \frac{x^2 y}{(x-y)^2}[g(x^{-1})-g(y^{-1})]\,, \nonumber\\
	I_2(x,y) &=& - \frac{x y}{2(x-y)}[f(x^{-1})-f(y^{-1})]\;,
\end{eqnarray}
 with the functions $f(x)$ and $g(x)$ in the range $x<1$, given by
\begin{eqnarray}
	f(x) = (\sin^{-1}\sqrt{x})^2\,, \qquad
	g(x) = \sqrt{x^{-1}-1}(\sin^{-1}\sqrt{x})\,.
\end{eqnarray}

The couplings of $h$ to the SM fermions and vector bosons {\it relative} to the SM Higgs couplings are given by
\begin{eqnarray}
	g_{hf\bar f} = \frac{\cos\alpha}{\cos\beta'}\, ,\qquad
	g_{hW^+W^-} = \cos\alpha+2\sin\alpha\frac{v_\Delta}{v}.
\label{coupSM}
\end{eqnarray}
From Eqs.~(\ref{mix1}) and (\ref{mix3}),
 we see in the limit $v_\Delta\ll v$, $\cos\alpha \simeq 1$, $\cos\beta' \simeq 1$,
 and hence %, $g_{hf\bar f} \simeq 1$, $g_{hW^+W^-} \simeq 1$.
 the couplings of $h$ to the SM fermions and vector bosons are almost identical to the SM case.
The couplings of $Z$ to the charged Higgs bosons in Eq.~(\ref{hZg}) are given by
\begin{eqnarray}
	g_{Z H^{+} H^{-}} = - \tan \theta_W\, , \qquad
	g_{Z H^{++} H^{--}} = 2\cot 2\theta_W \, .
\end{eqnarray}
For the scalar trilinear couplings, we have   
\begin{eqnarray}
	\tilde{g}_{h H^+H^-} = \frac{M_W}{g M_{H^{\pm}}^2} g_{h H^+H^-}\,, \qquad
	\tilde{g}_{h H^{++}H^{--}} = \frac{M_W}{ g M_{H^{\pm \pm}}^2} g_{h H^{++}H^{--}}\,,
\label{tri_coupling}
\end{eqnarray}
 with the following definitions in terms of the parameters of the scalar potential
 (up to ${\cal O}(v_\Delta^2)$)~\cite{Arhrib:2011uy}:
\begin{eqnarray}
	g_{hH^{+}H^{-}}  &=& \left[\left(\lambda_1 \cos^2\beta'+(\lambda_4+\lambda_5)
		\sin^2\beta'\right)v_\Delta+\sqrt 2 \lambda_5 \cos\beta'\sin\beta' v\right]
		\sin\alpha + \nonumber\\
		&& \left[\left(\lambda \sin^2\beta'+\lambda_4 \cos^2\beta'\right)v
		+ \sqrt 2\cos\beta'\sin\beta' \left(\frac{2M_\Delta^2}{v^2}+\lambda_4\right)v_\Delta\right]\cos\alpha\,,
\label{gh1}\\
	g_{hH^{++}H^{--}} &=& (\lambda_1+\lambda_2)v_\Delta \sin\alpha 
		+ (\lambda_4+\lambda_5)v\cos\alpha \,.
\label{gh2}
\end{eqnarray}
In the limit $v_\Delta\ll v$,
 Eqs.~(\ref{gh1}) and (\ref{gh2}) can be written as
\begin{eqnarray}
	g_{hH^{+}H^{-}}\simeq \lambda_4 v \,, \qquad
	g_{hH^{++}H^{--}} \simeq (\lambda_4+\lambda_5)v \,.	
\label{ghsimp}
\end{eqnarray}
Thus, the signs of the couplings $g_{hH^{+}H^{-}}$ and $g_{hH^{++}H^{--}}$, 
 and hence, those of the $H^\pm$ and $H^{\pm\pm}$ contributions to the amplitude in Eq.~(\ref{hgg})
 are fixed by the scalar couplings $\lambda_4$ and $\lambda_4+\lambda_5$, respectively.
The allowed parameter space by the vacuum stability and perturbativity conditions is shown in Fig.~\ref{scatter},
 and we can see that there is a small allowed region in $\lambda_4+\lambda_5<0$.

\begin{table}[t]
\begin{center}
\begin{tabular}{|c|cccc|}\hline
Amplitude & Fermions & $W$-Boson & $H^\pm$ & $H^{\pm\pm}$ \\
\hline \hline
$A^{\gamma \gamma}$ & $+$ & $-$ & $\lambda_4$ & $\lambda_4 + \lambda_5$ \\
$A^{Z \gamma}$ & $-$ & $+$ & $\lambda_4$ & $-(\lambda_4 + \lambda_5)$ \\
\hline
\end{tabular}
\end{center}
\caption{Signs of corresponding amplitudes.
Signs of the singly and doubly charged scalar contributions
 depend on signs of $\lambda_4$ and $\lambda_4 + \lambda_5$, respectively.}
\label{sign}
\end{table}

In the SM, the $W$-boson contributions to $h\to \gamma\gamma$ and $h\to Z\gamma$
 dominate over those from the SM fermions,
 while the signs of the corresponding amplitudes $A^{\gamma\gamma}_1$ and $A^{Z\gamma}_1$ are opposite
 as shown in Table~\ref{sign}.
The doubly charged scalar contribution usually dominates over the singly charged scalar contribution
 for both $h\to \gamma\gamma$ and $h\to Z\gamma$ amplitude
 because of the enhancement factor of four in Eqs.~(\ref{hgg}) and (\ref{hZg}),
 which corresponds to the squared electric charge of $H^{\pm\pm}$.
Since doubly charged scalar contributions are proportional to $\lambda_4+\lambda_5$
 with the opposite sign to the $W$-boson contribution,
 the both decay widths are enhanced for $\lambda_4+\lambda_5<0$.
For the same reason, the behavior reverses for $\lambda_4 + \lambda_5>0$.

In order to compare the model predictions for the signal strength with the SM value at the LHC,
 the partial decay widths of the processes $h\to \gamma\gamma,~Z\gamma $ can be expressed by
\begin{eqnarray}
	R_{\gamma\gamma (Z\gamma)}&=&
		\frac{\sigma(pp\to h \to \gamma\gamma (Z\gamma))}{\sigma_{\rm SM}(pp\to h \to \gamma\gamma (Z\gamma))}
	= \frac{\sigma(pp\to h)}{\sigma_{\rm SM}(pp\to h)}
		\frac{{\rm Br}(h\to \gamma\gamma (Z\gamma))}{{\rm Br}_{\rm SM}(h\to \gamma\gamma (Z\gamma))}\,,
\label{rggdef}
\end{eqnarray}
 where $\sigma(pp\to h)/\sigma_{\rm SM}(pp\to h) = \cos^2\alpha$
 with the mixing angle $\alpha$ given by Eq.~(\ref{mix3}).
Since $\cos\alpha \sim 1$ in the limit $v_\Delta \ll v$,
 the SM-like Higgs production rate is almost the same as that in the SM.
The branching ratios of all the Higgs decay channels are also the same as in the SM,
 except for $\gamma\gamma$ and $Z\gamma$ channels which may differ significantly,
 but their contribution to the total decay width remains negligible as in the SM.
Hence, for our numerical purposes,
 we can simply assume $R_{\gamma\gamma}$ defined in Eq.~(\ref{rggdef})
 to be the ratio of the partial decay widths for $h\to \gamma\gamma$ in the type-II seesaw model and in the SM.

\begin{figure}[t]
\begin{center}
	\includegraphics[height=7cm]{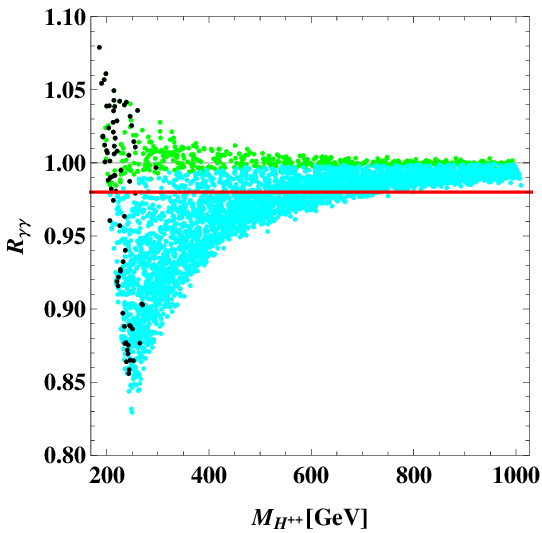} \hspace{5mm}
	\includegraphics[height=7cm]{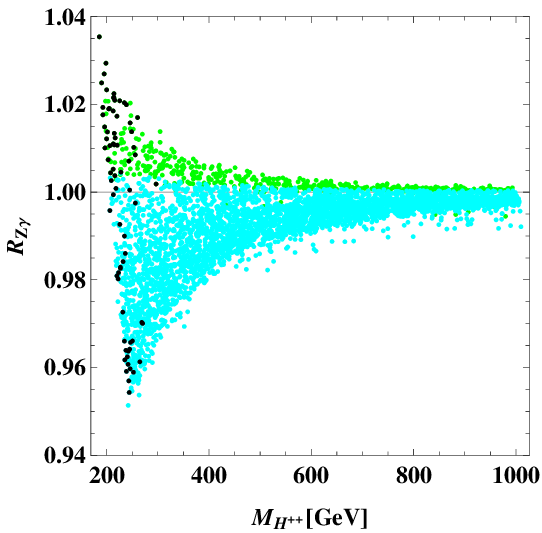}
\end{center}
\caption{$R_{\gamma\gamma}$ (left) and $R_{Z\gamma}$ (right) versus $M_{H^{\pm\pm}}$.
The green and cyan dots correspond to $\lambda_4 + \lambda_5 < 0$ and $\lambda_4 + \lambda_5 >0$, respectively.
The black dots satisfy $|\delta m_h^2|<M_h^2$.
The red line in the left panel shows the experimental lower bound.}
\label{Rgg}
\end{figure}

Figure \ref{Rgg} shows $R_{\gamma\gamma}$ (left) and $R_{Z\gamma}$ (right) versus $M_{H^{\pm\pm}}$.
The green and cyan dots correspond to
 $\lambda_4 + \lambda_5 < 0$ and $\lambda_4 + \lambda_5 >0$, respectively.
The black dots satisfy $|\delta m_h^2|<M_h^2$.
The red line in the left panel shows the lower bound on the current signal strength,
 which is obtained as $R_{\gamma \gamma}=1.16^{+0.20}_{-0.18}$
 by the combined analysis of ATLAS and CMS results~\cite{Hto2gamma}.
Our result of $R_{\gamma \gamma} \lesssim 1.1$
 needs more than 10 \% precision to see the deviation,
 while the relative uncertainty on the $R_{\gamma \gamma}$ is 0.1 for the combined Higgs analysis
 by the LHC experiment at 14\,TeV with $3000\,{\rm fb}^{-1}$ of integrated luminosity~\cite{ATLAS_gg}.
There is no useful experimental bound for $R_{Z \gamma}$ at present~\cite{Aad:2014fia}.
The expected measured signal for $h \to Z\gamma \to \ell \ell \gamma$ is
 $1.00^{+0.25}_{-0.26}({\rm stat.})^{+0.17}_{-0.15}({\rm sys.})$
 by the 14\,TeV LHC experiment with $3000\,{\rm fb}^{-1}$~\cite{ATLAS_Zg}.

In the $\lambda_4 + \lambda_5 < 0$ case,
 both $R_{\gamma\gamma}$ and $R_{Z\gamma}$ are larger than unity
 with $M_{H^{\pm\pm}} < M_\Delta$.
The behavior reverses for $\lambda_4 + \lambda_5>0$:
 in the $\lambda_4 + \lambda_5 > 0$ case,
 both $R_{\gamma\gamma}$ and $R_{Z\gamma}$ are smaller than unity
 with $M_{H^{\pm\pm}} > M_\Delta$.
Although $R_{\gamma\gamma}$ can be enhanced by both $H^{\pm}$ and $H^{\pm\pm}$ contributions
 for $\lambda_4<0$,
 there is no parameter space in $\lambda_4<0$ region for $M_\Delta \leq 1\,{\rm TeV}$
 [see the right panel in Fig.~\ref{scatter}].
This result comes from the vacuum stability conditions,
 and we note that $R_{\gamma\gamma}$ is not strongly enhanced
 compared to the literature; for example Ref.~\cite{Arhrib:2011vc,Akeroyd:2012ms}.

\begin{figure}[t]
\begin{center}
	\includegraphics[height=8cm]{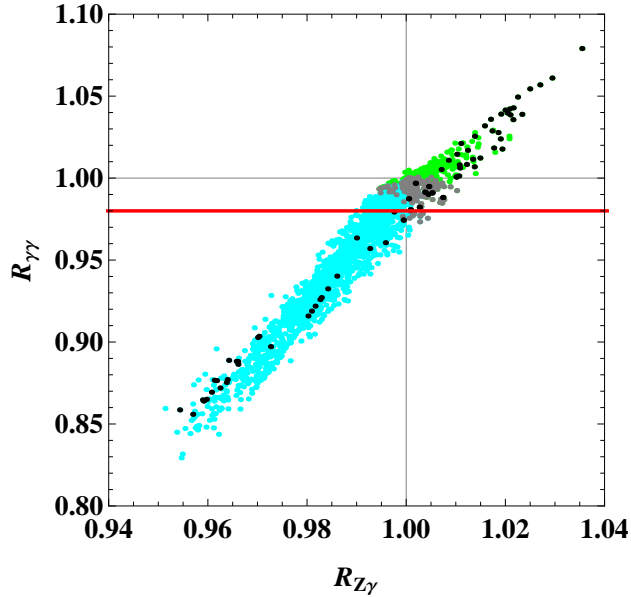}
\end{center}
\caption{The relation between $R_{\gamma\gamma}$ and $R_{Z\gamma}$.
The green, gray, and cyan dots correspond to
 $\lambda_4 + \lambda_5 < -0.05$, $-0.05 < \lambda_4 + \lambda_5 < 0.05$,
 and $\lambda_4 + \lambda_5 >0.05$, respectively.
The black dots satisfy $|\delta m_h^2|<M_h^2$.
The red line shows the experimental lower bound.}
\label{ggZg}
\end{figure}

In the $\lambda_4 + \lambda_5 \simeq 0$ case,
 contributions from $H^{\pm\pm}$ to the decay rate vanish,
 while contributions from $H^{\pm}$ can be seen clearly.
We can see from Table~\ref{sign}
 that there is an anti-correlation between $R_{\gamma\gamma}$ and $R_{Z\gamma}$
 for $\lambda_4 + \lambda_5 \simeq 0$.
To see this,
 we show the relation between $R_{\gamma\gamma}$ and $R_{Z\gamma}$ in Fig.~\ref{ggZg}.
The gray dots correspond to $-0.05 < \lambda_4 + \lambda_5 < 0.05$ with $\lambda_4>0$,
 and they lie in the $R_{\gamma\gamma}<1$ and $R_{Z\gamma}>1$ region.
Since there is no allowed parameter space in $-0.05 < \lambda_4 + \lambda_5 < 0.05$ with $\lambda_4<0$,
 the model cannot realize $R_{\gamma\gamma}>1$ and $R_{Z\gamma}<1$ at the same time.

%%%%%%%%%%%%%%%%%%%%%%%%%%%%%%%%%%%%%%%%%%%
\section{Conclusion} \label{sec:conclusion}
%%%%%%%%%%%%%%%%%%%%%%%%%%%%%%%%%%%%%%%%%%%

We have studied the vacuum stability and perturbativity conditions in the minimal type-II seesaw model.
Their conditions give characteristic constraints between model parameters as in Fig.~\ref{scatter}.
The vacuum stability can be realized by sufficiently large $|\lambda_4|$ or $\lambda_6$,
 which leads to large Higgs mass corrections.
To realize the naturalness condition ($\delta m_h^2 < M_h^2 = (125\,{\rm GeV})^2$),
 we have found that heavy Higgs masses should be lower than $350\,{\rm GeV}$.
Below this bound, the minimal type-II seesaw model can be testable by the LHC Run-II results.
Due to the triplet scalar field,
 branching ratios of the Higgs decay for $h\to \gamma \gamma, Z\gamma$
 are different from the standard model case.
They strongly depend on the sign of $\lambda_4+\lambda_5$,
 and there is an anti-correlation between $R_{\gamma \gamma}$ and $R_{Z \gamma}$
 for $\lambda_4+\lambda_5 \simeq 0$ with $\lambda_4>0$.
From the recent ATLAS and CMS combined analysis for the signal strength of $h\to \gamma \gamma$,
 we have also found that a large parameter region is to be excluded.
Our result of $R_{\gamma \gamma} \lesssim 1.1$
 needs more than 10 \% precision to see the deviation,
 while the relative uncertainty on the $R_{\gamma \gamma}$ is 0.1 for the combined Higgs analysis
 by the LHC experiment at 14\,TeV with $3000\,{\rm fb}^{-1}$ of integrated luminosity.

%%%%%%%%%%%%%%%%%%%%%%%%%%%%%%%%%%%%%%%%%%%
\section*{Acknowledgment} \label{Acknowledgement}
%%%%%%%%%%%%%%%%%%%%%%%%%%%%%%%%%%%%%%%%%%%
N. O. would like to thank the Particle Physics Theory Group of Shimane University
  for hospitality during his visit.
This work is partially supported by Scientific Grants
  by the Ministry of Education, Culture, Sports, Science and Technology (Nos. 24540272, 26247038, and 15H01037)
  and the United States Department of Energy (DE-SC 0013680).
The work of Y. Y. is supported
  by Research Fellowships of the Japan Society for the Promotion of Science for Young Scientists
  (Grants No. 26$\cdot$2428).

%%%%%%%%%%%%%%%%%%%%%%%%%%%%%%%%%%%%%%%%%%%
\section*{Appendix}
%%%%%%%%%%%%%%%%%%%%%%%%%%%%%%%%%%%%%%%%%%%
\appendix
%%%%%%%%%%%%%%%%%%%%%%%%%%%%%%%%%%%%%%%%%%%
\section*{The beta functions in the minimal type-II seesaw model} \label{app:RGE}
%%%%%%%%%%%%%%%%%%%%%%%%%%%%%%%%%%%%%%%%%%%

The one-loop beta functions for the minimal type-II seesaw model are given by
\begin{eqnarray}
16\pi^2 \beta_{g_Y} &=& \frac{47}{6} g_Y^3,\qquad 
16\pi^2 \beta_{g_2} = - \frac{5}{2} g_2^3 ,\qquad
16\pi^2 \beta_{g_3} = -7 g_3^3 \\
16\pi^2 \beta_{y_t} &=&
	y_t \left[ \frac{9}{2} y_t^2 - \left( \frac{17}{20} g_1^2 + \frac{9}{4} g_2^2 + 8 g_3^2 \right) \right], \\
16\pi^2 \beta_\lambda &=&
	\lambda \left[ 12\lambda - \left( \frac{9}{5}g_1^2 + 9g_2^2 \right) + 12y_t^2 \right]
	+ \frac{9}{4} \left( \frac{3}{25}g_1^4 + \frac{2}{5}g_1^2g_2^2 + g_2^4 \right) \nonumber \\
	&& + 6\lambda_4^2 + 4\lambda_5^2 - 12y_t^4,
\label{beta} \\
16\pi^2 \beta_{\lambda_1} &=&
	\lambda_1 \left[ 14\lambda_1 + 4 \lambda_2
	- \left( \frac{36}{5} g_1^2 + 24g_2^2 \right)
	+ 4 {\rm tr} \left[{\bf S}_\Delta \right] \right]
	+ \frac{108}{25}g_1^4 + \frac{72}{5}g_1^2g_2^2 + 18g_2^4 \nonumber\\
	&& + 2 \lambda_2^2 + 4 \lambda_4^2 + 4\lambda_5^2
	- 8 {\rm tr} \left[{\bf S}_\Delta^2 \right], \\
16\pi^2 \beta_{\lambda_2} &=&
	\lambda_2 \left[ 12 \lambda_1 + 3 \lambda_2 
	- \left( \frac{36}{5}g_1^2 + 24g_2^2 \right)
	+ 4 {\rm tr} \left[{\bf S}_\Delta  \right] \right]
	- \frac{144}{5}g_1^2g_2^2 + 12g_2^4 \nonumber\\
	&& - 8 \lambda_5^2 + 8 {\rm tr} \left[{\bf S}_\Delta^2 \right],  \\
16\pi^2\ \beta_{\lambda_4} &=&
	\lambda_4 \left[ 6 \lambda + 8 \lambda_1 + 2 \lambda_2 + 4\lambda_4
	- \left( \frac{9}{2}g_1^2 + \frac{33}{2}g_2^2 \right) 
	+ 6 y_t^2 + 2 {\rm tr} \left[{\bf S}_\Delta \right] \right] \nonumber\\
	&& + \frac{27}{25}g_1^4 + 6g_2^4
	+ 8 \lambda_5^2 - 4 {\rm tr}\left[ {\bf S}_\Delta^2 \right], \\
16\pi^2 \beta_{\lambda_5} &=&
	\lambda_5 \left[ 2 \lambda + 2\lambda_1 - 2\lambda_2 + 8 \lambda_4
	- \left( \frac{9}{2}g_1^2 + \frac{33}{2}g_2^2 \right)
	+ 6 y_t^2 + 2 {\rm tr}\left[{\bf S}_\Delta \right] \right]
	- \frac{18}{5}g_1^2g_2^2 \nonumber\\
	&& + 4 {\rm tr}\left[{\bf S}_\Delta^2 \right],
\end{eqnarray}
 where we define ${\bf S}_\Delta=Y_\Delta^\dagger Y_\Delta $
 and its beta function is given by
\begin{eqnarray}
16\pi^2 \beta_{{\bf S}_\Delta} =
	{\bf S}_\Delta \left[ 6\, {\bf S}_\Delta - 3 \left( \frac{3}{5} g_1^2 + 3 g_2^2 \right)
	+ 2 {\rm tr}[{\bf S}_\Delta] \right] .
\end{eqnarray}

% \newpage
%%%%%%%%%%%%%%%%%   bibliography   %%%%%%%%%%%%%%%%%%%

\end{document}